\documentstyle[aps,pre,preprint]{revtex}
\begin{document}

\draft
\title{Lonely adatoms in space}
\author{Joachim Krug}
\address{Fachbereich Physik, Universit\"at Duisburg-Essen,
45117 Essen, Germany
}
\date{\today}
\maketitle
\begin{abstract}
There is a close relation between the problems of second layer
nucleation in epitaxial crystal growth and chemical surface reactions, such as hydrogen recombination, 
on interstellar dust grains. In both cases standard rate equation analysis has been found
to fail because the process takes place in a confined geometry. Using scaling arguments
developed in the context of second layer nucleation, 
I present a simple derivation of the hydrogen recombination rate for small and large grains.
I clarify the reasons for the failure of rate equations for small grains, and point out a logarithmic
correction to the reaction rate when the reaction is limited by the desorption of hydrogen atoms
(the second order reaction regime).
\end{abstract}



Rate equations are a widely used tool for the mathematical description
of kinetic phenomena. Despite successful applications in many
different areas of science, however, there are 
instances where their predictions
are known to be qualitatively incorrect. A
prominent example are diffusion-controlled reactions in low dimensions
\cite{Redner97}. 
The purpose of this Rapid Communication 
is to point out the close relation between
two kinetic problems, second layer nucleation in epitaxial crystal
growth \cite{Heinrichs00,Krug00a,Krug00b,Castellano01} and chemical reactions 
on dust grain surfaces in interstellar
space \cite{Green01,Biham01,Biham02,Williams02}, in which deviations from
rate equation theory arise because the processes of interest
take place in a spatially confined geometry. 
 
In both cases one is concerned with the
rate of a reaction confined to a finite surface region,
onto which particles impinge and from which they leave again
if no reaction takes place. In the case of second layer
nucleation the confining region is the top terrace of a two-dimensional island,
bounded by a step edge, while in the astrophysical context it is
the (closed) surface of a small grain. In the first case the atoms leave
the island by crossing the step, while in the second case they desorb
from the grain surface\footnote{At high temperatures desorption plays
a role also in epitaxial growth; see \cite{Jensen97} for an analysis
of two-dimensional nucleation in the presence of desorption.}. 

The two problems also differ in that in second layer nucleation
only the first nucleation event on the island is relevant -- all 
atoms deposited after this event merely contribute to the growth of the
second layer nucleus -- whereas in the problem of grain surface reactions
the reactants are usually assumed to desorb from the grain once
they have formed, so that a steady state with a constant
production rate is established. 
In this sense, second layer nucleation is a first passage problem 
\cite{Castellano01}, while grain surface kinetics deals with stationary
processes. 

For both problems it has been found 
that standard rate equation analysis fails when the mean number of 
particles present on the surface becomes small compared to unity
(the \emph{lonely adatom} limit of Ref.\cite{Krug00a}).
In this fluctuation-dominated
regime the reaction becomes a rare event which has to 
be treated statistically.
Here I show how the time scale analysis developed in the
context of second layer nucleation 
\cite{Krug00a,Krug00b} can be applied
to the problem of hydrogen recombination, one of 
the most important grain surface reactions of astrophysical 
interest. In this reaction two hydrogen atoms meet to form a H$_2$ molecule, 
in analogy to the formation of a second layer nucleus
in the case where dimers are stable (irreversible 
nucleation, critical nucleus size $i^\ast = 1$).
I provide a simple derivation
of the recombination rate in all three reaction regimes. I then discuss the 
reasons for the breakdown of the rate equation theory for small grains, 
and I point out the
existence of a logarithmic correction to the recombination rate in the second
order reaction regime, which was overlooked so far in the astrophysical context.

The notation follows that of \cite{Biham02}. Atoms impinge on the grain surface
at rate $f$ per adsorption site, they hop at rate $a$ and desorb at rate $W$. The number
of adsorption sites on the grain is $S$, and the quantities
$F = fS$ and $A = a/S$ refer to the whole grain. To make contact
with the notation of \cite{Krug00a,Krug00b} we introduce the lifetime
$\tau$ of an atom on the grain,
\begin{equation}
\label{tau}
\tau = 1/W,
\end{equation} 
the diffusion time $\tau_D$ required for an atom to explore the whole grain,
\begin{equation}
\label{tauD}
\tau_D \approx S/a = 1/A,
\end{equation}
and the mean time interval $\Delta t$ between the arrival of two atoms
on the grain,
\begin{equation}
\label{Deltat}
\Delta t = 1/F.
\end{equation}
The estimate (\ref{tauD}) ignores a logarithmic correction arising
from the distinction between the total 
number of sites visited by an adatom during its lifetime, 
${\cal{N}}_{\mathrm{all}}(\tau)$, 
and the number ${\cal{N}}_{\mathrm{dis}}(\tau)$ of 
\emph{distinct} sites visited; we will return to this point towards the end
of the paper.

Following \cite{Biham02} we introduce the dimensionless quantities 
\begin{equation}
\label{svisit}
s_{\mathrm{visit}} = a/W = S (\tau/\tau_D)
\end{equation}
and
\begin{equation}
\label{svacant}
s_{\mathrm{vacant}} = W/f = S(\Delta t/\tau).
\end{equation}
For large grains ($S > s_{\mathrm{visit}}$ and $S > s_{\mathrm{vacant}}$, respectively)
$s_{\mathrm{visit}}$ is the number of sites visited by an atom during its lifetime, 
while $s_{\mathrm{vacant}}$ is the number of vacant sites surrounding a site occupied
by an atom.
If $\tau > \tau_D$ the atom has time to eplore the whole grain, so that 
the number of sites visited becomes equal to 
$S$, while for $\tau < \Delta t$ the grain is typically occupied
by zero or one atoms, so that an atom is surrounded by $S$ vacant sites; 
note that the mean number of atoms on the grain is equal to 
\begin{equation}
\label{nbar}
\bar n = \tau/\Delta t = F/W.
\end{equation}
The relative ordering of the
three dimensionless numbers 
$S$, $s_{\mathrm{visit}}$ and $s_{\mathrm{vacant}}$ defines six different regimes.
It turns out, however, that the behavior of the reaction rate in a given regime depends only
on which of the three numbers is the smallest. Thus we are left with three different regimes,
which will be discussed in the following. 

The fluctuation-dominated (lonely adatom) regime is characterized by the condition
\cite{Biham02}
\begin{equation}
\label{small}
S \ll \min[s_{\mathrm{visit}}, s_{\mathrm{vacant}}].
\end{equation}
According to (\ref{svisit},\ref{svacant}) this is equivalent to 
(i) $\tau \ll \Delta t$ and (ii) $\tau \gg \tau_D$. The first
condition simply implies that the mean number of atoms on the grain (\ref{nbar})
is small compared to unity. In this limit, the probability that  
a freshly deposited atom participates in a recombination
event is equal to the probability $P_1$ that a second atom is already 
present on the grain, times the \emph{encounter probability} $p_{\mathrm{enc}}$
that the two atoms, once both present on the grain, 
will meet before one of them redesorbs.
The first probability is given by \cite{Krug00a}
$P_1 = \tau/(\tau + \Delta t) \approx \tau/\Delta t = F/W$. Multiplying
this with the number of atoms $F = 1/\Delta t$ arriving on the grain
per unit time, we find that the total recombination rate 
(the number of molecules produced by the grain per unit time)
is given by
\begin{equation}
\label{Rgen}
R = \frac{\tau}{(\Delta t)^2} \; p_{\mathrm{enc}}
\end{equation}
whenever $\bar n \ll 1$. 
The second condition $\tau \gg \tau_D$ implied by (\ref{small}) means that,
if two atoms are present simultaneously on the grain, they will
meet with unit probability, because each atom spends enough
time on the grain to visit all sites many times. Thus we have 
$p_{\mathrm{enc}} = 1$ and 
\begin{equation}
\label{R}
R = R_{\mathrm{lonely}} = \tau/(\Delta t)^2 = F^2/W.
\end{equation}
In the small grain limit where (\ref{small}) is satisfied, this
expression is \emph{exact}. 
The recombination efficiency defined by $\eta = R/(F/2)$ 
is then equal to 
$ 2 F/W$, in agreement with the result obtained in \cite{Biham02}
from the full solution of the master equation. 
In second layer nucleation, the 
lonely adatom regime is realized under typical experimental conditions 
when $i^\ast = 1$ \cite{Krug00a}.   

In the remaining two regimes the finite grain size is irrelevant. These regimes can therefore
be derived from standard rate equations \cite{Biham98}. According to the scaling of the
reaction rate with hydrogen flux, these two regimes will be referred to as the \emph{first} and
\emph{second order} regimes (note that the lonely adatom regime is also of second order in this
sense, see (\ref{R})). 

The first order regime is characterized by the condition    
\begin{equation}
\label{first}
s_{\mathrm{vacant}} \ll \min[S, s_{\mathrm{visit}}].
\end{equation}
In this regime the mean number of atoms on the grain is large compared to unity
($\tau \gg \Delta t$), and the number of sites visited by an atom during its lifetime
($= \min[S, s_{\mathrm{visit}}]$) 
is large compared to the number of vacant sites
surrounding each atom. Hence each freshly deposited atom
will certainly find a partner. In this case the recombination
rate is 
\begin{equation}
\label{R1}
R = R_1 = F/2
\end{equation} 
and the recombination efficiency is equal to
unity. In the context of second layer nucleation, this regime corresponds to the 
nonstationary scaling regime I 
of \cite{Heinrichs00,Krug00b}, where the energy barriers at the step edges are so strong that
$\tau \gg \Delta t$, and nucleation occurs as soon as a second atom is deposited onto the island. 

It is easy to derive an interpolation formula connecting (\ref{R}) and (\ref{R1}) under
the assumption that the diffusion of the reactants, as well as the desorption of the hydrogen
molecule, is very fast. Then the only possible values for the number of atoms on the grain
is 0 or 1, since recombination occurs instantaneously as soon as a second atom is deposited.
The probability $P_1$ for having one atom on the grain satisfies the master equation
\begin{equation}
\label{P1}
d P_1/dt = F P_0 - (F + W) P_1 = F (1 - P_1) - (F + W) P_1
\end{equation}
with the stationary solution $P_1 = F/(W + 2 F)$. Since $p_{\mathrm{enc}} = 1$,
the recombination rate is 
\begin{equation}
\label{Rinter}
R = F P_1 = \frac{F^2}{W + 2 F}.
\end{equation}

The remaining, second order regime is characterized by the condition  
\begin{equation}
\label{secondcond}
s_{\mathrm{visit}} \ll \min[S, s_{\mathrm{vacant}}].
\end{equation}
The condition $s_{\mathrm{visit}} \ll S$ (i.e., $\tau \ll \tau_D$) has no clear analogue
in the context of second layer nucleation, because there the lifetime of the atom is always
at least as large as the diffusion time (the atom has to reach the island edge before 
it can leave the island). 
It is useful to distinguish the cases $S \gg s_{\mathrm{vacant}}$
(large grains) and $S \ll s_{\mathrm{vacant}}$ (small grains). In the first
case many atoms are present on the grain simultaneously. The encounter
probability for a freshly deposited atom to find a partner during its lifetime 
is equal to $p_{\mathrm{enc}} = s_{\mathrm{visit}}/s_{\mathrm{vacant}}$. Thus
the recombination rate can be estimated as
\begin{equation}
\label{second1}
R = R_2 = F p_{\mathrm{enc}} = F \frac{\tau^2}{\tau_D \Delta t} = \frac{f^2 S a}{W^2}
\end{equation}
and the recombination efficiency is $\eta = 2 a f/W^2$. For
small grains the probability that a freshly deposited atom will undergo a 
reaction is the product of the probability $P_1 = F/W$ that a second atom
is present on the grain when it arrives, and the encounter probability, which
now reads $p_{\mathrm{enc}} = s_{\mathrm{visit}}/S$ (note that the second atom
occupies any site of the grain with equal probability). The recombination rate
then becomes 
\begin{equation}
\label{second1b}
R = \frac{F^2}{W} p_{\mathrm{enc}} = \frac{f^2 S a}{W^2}, 
\end{equation}
which is \emph{identical} to (\ref{second1}). Thus the second order result hold irrespective
of whether the number of atoms on the grain is large compared to unity or not.
This is similar to certain regimes of second layer nucleation with critical nucleus
size $i^\ast > 1$ \cite{Krug00b}.  

The last observation shows that the failure of rate equation theory cannot generally
be attributed to the fact that the number of reactants on the grain is small compared
to unity. To clarify this issue, it is instructive to consider the amount by which
the reaction rate is overestimated by the rate equation expressions when the condition
(\ref{small}) is satisfied. Two cases have to be distinguished. For 
$s_{\mathrm{vacant}} \ll s_{\mathrm{visit}}$ we have to compare (\ref{R})
to $R_1$, which yields 
\begin{equation}
\label{ratio1}
R_{\mathrm{lonely}}/R_1 = F/W = \tau/\Delta t.
\end{equation}
This is precisely the mean number of atoms $\bar n$ on the grain. Thus in this case
it is justified to say that the rate equations fail because $\bar n \ll 1$. On the 
other hand, for $s_{\mathrm{visit}} \ll s_{\mathrm{vacant}}$ we have    
\begin{equation}
\label{ratio2}
R_{\mathrm{lonely}}/R_2 = S W/a = S/s_{\mathrm{visit}}.
\end{equation}
This ratio is not related to the mean number of atoms on the grain. Instead, it
is equal to the ratio between the number of distinct sites visited by an adatom
during its lifetime (${\cal{N}}_{\mathrm{dis}} = S$) and the number of all sites
visited (${\cal{N}}_{\mathrm{all}} = a/W = s_{\mathrm{visit}}$). Castellano and
Politi have emphasized that it is the distinction between ${\cal{N}}_{\mathrm{dis}}$ and
${\cal{N}}_{\mathrm{all}}$ which is responsible for the failure of rate equation theory   
in the case of second layer nucleation \cite{Castellano01}. 

In the second order regime this distinction implies a logarithmic correction to the
expression (\ref{second1}). The number of distinct sites
visited by a two-dimensional random walk during time $\tau$ is given by  
${\cal{N}}_{\mathrm{dis}}(\tau) \approx \pi a \tau/\ln(a \tau)$
\cite{Itzykson89}. The diffusion time should therefore be determined more accurately
by ${\cal{N}}_{\mathrm{dis}}(\tau_D) = S$, and the expression (\ref{svisit}) has to 
be replaced by  
$s_{\mathrm{visit}} = {\cal{N}}_{\mathrm{dis}}(\tau)$. Using this in the derivation of 
the reaction rate in the second order regime we find
\begin{equation}
\label{second2}
R_2 = \frac{f^2 S}{W} s_{\mathrm{visit}} \approx \frac{\pi f^2 S a}{W^2 \, \ln(a/W)}.
\end{equation}
The experimental estimates for the activation energies for H diffusion and desorption
on amorphous carbon surfaces quoted in 
\cite{Biham02} show that the logarithmic factor in (\ref{second2}) is $\ln(a/W) \approx 15$
at a temperature of 10 K. 
 
To go beyond the order of magnitude estimate (\ref{second2}) would require a detailed analysis
of the random walk problem on the grain surface along the lines of \cite{Castellano01}.
The master equation analysis of \cite{Green01,Biham01,Biham02} is not sufficient for this purpose,
because it keeps track only of the total number of atoms on the grain, thus ignoring the spatial
aspects of the encounters between the atoms. As a first step in this direction, I present
here a simple model calculation of the encounter probability $p_{\mathrm{enc}}$ 
for the case of small grains (in the sense of $\bar n \ll 1$),
which allows
to interpolate between the expressions (\ref{R}) and (\ref{second2}); a similar approach has
previously been applied in the context of second layer nucleation \cite{Michely03}.

We take the grain surface to be a circular disk of radius $L$. One of the reacting atoms
is assumed to be immobile, and is placed at the center of the large disk in the form of a small
disk of radius $l$. In this way the problem becomes radially symmetric.
The deposition, diffusion and desorption of the second atom is described by 
the stationary diffusion equation 
\begin{equation}
\label{diffusion}
D \left( \frac{\partial^2}{\partial r^2} + \frac{\partial}{\partial r} \right) n + \tilde f - W n = 0
\end{equation}
where $n(r)$ is the probability density for the second atom 
in polar coordinates, $\tilde f$ is the impingement flux per unit area,
and the diffusion coefficient is
$D = a l^2$. The boundary conditions
for (\ref{diffusion}) are reflecting at the outer boundary, $\partial n/\partial r = 0$
at $r=L$, and absorbing at the inner boundary ($n(r = l) = 0$). The encounter probability
is then obtained as the ratio of the diffusion flux incident onto the inner boundary to
the total deposition flux incident onto the surface,
\begin{equation}
\label{penc}
p_{\mathrm{enc}} = \frac{2 \pi  l D \; \partial n/\partial r \vert_{r=l}}{\tilde f \pi L^2}.
\end{equation}
The solution to this diffusion problem can be found in the thin film growth
literature \cite{Stowell72}.
In the limit where the atom radius $l$ becomes the smallest length in the problem,
the encounter probability takes the form
\begin{equation}
\label{penc2}
p_{\mathrm{enc}} = \frac{4 a}{S W} \left( \ln(a/W) + 
2 K_1(\sqrt{S W/a})/I_1(\sqrt{S W/a})\right)^{-1},
\end{equation}
where $I_1$ and $K_1$ are incomplete Bessel functions. For small arguments 
$K_1(x)/I_1(x) \approx 2/x^2$ and $p_{\mathrm{enc}} \to 1$, while for large
arguments $K_1(x)/I_1(x) \to 0$ and (\ref{penc2}) reduces to 
$p_{\mathrm{enc}} \approx 4 a /(S W \ln (a/W))$. Inserting the second result into 
(\ref{Rgen}) we obtain an expression identical to (\ref{second2}), apart from the
factor $\pi$ which is replaced by a factor 4. 

To summarize, I have shown in this paper how recent results from the theory
of second layer nucleation can be applied to the closely related problem of 
hydrogen recombination on dust grain surfaces. 
The expressions (\ref{R}), (\ref{R1}) and (\ref{second1}) for
the recombination rate have been obtained before \cite{Biham02}, but the present
derivation seems more straightforward. 
The analysis shows that the failure of the rate equation
approach is not always related to the fact that the mean number of atoms $\bar n$ on
the grain is small compared to unity.  
The logarithmic factor in the second order
regime, Eq.(\ref{second2}), and the interpolation formulae (\ref{Rinter}) and
(\ref{penc2}) are new. The derivation of the encounter probability 
for realistic grain geometries using the methods of \cite{Castellano01}
seems like a promising problem for future studies.

\noindent
\emph{Acknowledgement:} I would like to thank Ofer Biham for useful correspondence.

\end{document}